# Mechanical properties of lateral transition metal dichalcogenide heterostructures

Sadegh Imani Yengejeh,[1] William Wen,[1] Yun Wang[1*]

1. Centre for Clean Environment and Energy, School of Environment and science, Griffith University, Gold Coast campus, Southport 4222, Australia

* Corresponding author: yun.wang@griffith.edu.au

## ABSTRACT

Transition metal dichalcogenide (TMD) monolayers attract great attention due to their specific structural, electronic and mechanical properties. The formation of their lateral heterostructures allows a new degree of flexibility in engineering electronic and optoelectronic devices. However, the mechanical properties of the lateral heterostructures are rarely investigated. In this study, a comparative investigation on the mechanical characteristics of 1H, 1T' and 1H/1T' heterostructure phases of different TMD monolayers including molybdenum disulfide ($MoS_2$) molybdenum diselenide ($MoSe_2$), Tungsten disulfide ($WS_2$), and Tungsten diselenide ($WSe_2$) was conducted by means of density functional theory (DFT) calculations. Our results indicate that the lateral heterostructures have a relatively weak mechanical strength for all the TMD monolayers. The significant correlation between the mechanical properties of the TMD monolayers and their structural phases can be used to tune their stiffness of the materials. Our findings, therefore, suggest a novel strategy to manipulate the mechanical characteristics of TMDs by engineering their structural phases for their practical applications.

# I. INTRODUCTION

With superb electronic and mechanical properties, two-dimensional (2D) ultrathin functional materials have shown great promise in a wide range of applications [1-3]. Over the past few years, layer-structured transition metal dichalcogenides (TMDs) such as Tungsten disulfide ($WS_2$), Tungsten diselenide ($WSe_2$), molybdenum disulfide ($MoS_2$), and molybdenum diselenide ($MoSe_2$) have attracted increasing attention due to their significant characteristics and physical properties including large exciton binding energy [4], band gap transition [5], and abundance of multiexcitons [6]. Due to these remarkable properties, an increasing interest and also recent progress led to a wide range of applications of TMDs including energy storage [7], sensors [8], and batteries [9] which demonstrate the high capacity of the materials in industry sectors. Basically, TMD monolayers can be described as a sandwich type of structure (X-TM-X) in which metal atoms (e.g. Mo and W) are located in between two layers of chalcogen atom (e.g. S and Se). 2D TMD monolayers can exist in various polymorphs, wherein subtle structural changes can dramatically affect the electrical characteristics. 1H $MoS_2$ is found as the semiconducting and thermodynamically favored phase, which is described by two X−TM−X layers built from edge-sharing $TMX_6$ trigonal prisms. In contrast, the metallic 1T, 1T' and 1T'' polymorphs are described by a single X−TM−X layer composed of edge-sharing $TMX_6$ octahedra, which are not naturally found in bulk due to less thermodynamic stability [10, 11].

To broaden the application of 2D materials, efforts have been devoted to the development of atomic scale heterostructures [12-16]. Individually, atomically thin layers arising from the successful exfoliation of 2D monolayers can serve as building blocks for promising hybrid materials, where 2D materials are either vertically layer-by-layer stacked by van der Waals (vdW) forces, or growth together seamlessly in-plane to form lateral heterojunctions by covalent bonds. The lateral heterojunctions have recently been paid more attention since they can greatly affect the physicochemical properties of 2D materials [17-19]. In contrast to their extensively explored electrical and optical characteristics, mechanical behaviors of these 2D lateral heterostructures have not been well characterized. Our recent results suggest that the structural phase can greatly affect their mechanical properties of TMD bulk systems [20]. Thus, the understanding of the lateral TMD heterostructures becomes imperative.

Recently, the classic molecular dynamics (MD) simulation based on the valence force field has been conducted to understand their buckling behavior[21, 22]. However, the atomic understanding by virtual of the first-principles method is still rare. The aim of this study is to understand the mechanical characteristics of the 1H/1T' laterally heterostructured TMD monolayers. Four TMDs including $MoS_2$, $MoSe_2$, $WS_2$ and $WSe_2$ were selected as the model systems because they are the most studied TMD systems with the wider applications in practical. As the reference, the properties of the 1H and 1T' TMD monolayers were also investigated. Our results reveal that the formation of the laterally heterostructured TMD monolayers reduces their stiffness. Such change of the mechanical properties may be beneficial to the strain engineering for the application in energy and environment areas.

## II. COMPUTATIONAL DETAILS

In this study, all DFT computations were conducted using the Vienna ab initio simulation package (VASP) [23-25]. The generalized gradient approximation (GGA) with the format of Perdew-Burke-Ernzehof (PBE) was used for the exchange-correlation functional with the DFT-D3 dispersion corrections [26-28]. In our calculation, a plane-wave basis set with a cut-off kinetic energy of 520 eV was employed to expand the smooth part of wave function. And gamma point centered ($18 \times 18 \times 1$), ($9 \times 9 \times 1$), and ($1 \times 3 \times 1$) k-point grids for 1H, 1T', and 1H/1T' phases of TMDs were employed in our simulation, respectively. Before the calculation of mechanic properties, both the lattice constants and the atomic coordinates were optimized.

All the atoms were allowed to relax until the Hellmann-Feynman forces were smaller than 0.02 eV/Å, and the convergence criterion for the self-consistent electronic optimization loop was set to $1 \times 10^{-5}$ eV.

To investigate the elastic constants of the TMDs monolayers according to generalized Hooke's law, six finite distortions of the lattice were performed to derive the elastic constants from the strain-stress relationship [29]. The in-plane stiffness tensors $\boldsymbol{C_{ij}}$ in N/m were obtained by multiplying the three-dimensional stiffness tensor with the height of the unit cell. The selection of the height is validated through the comparison with the reported results of the know 1H TMDs.

## III. RESULTS AND DISCUSSION

The atomic configurations of TMDs with different phases (1H, 1T', and 1H/1T') are illustrated in **Fig. 1**. The optimized structures are provided in the **Supporting Information**. These three phases are selected because the lateral 1H/1T' $MoS_2$ heterostructure has been experimentally synthesized, which shows promising applications in the energy area [30]. As the reference, the 1H and 1T' phases are also studied for understanding the impact of the lateral heterostructure on their structural, electronic and mechanical characteristics.

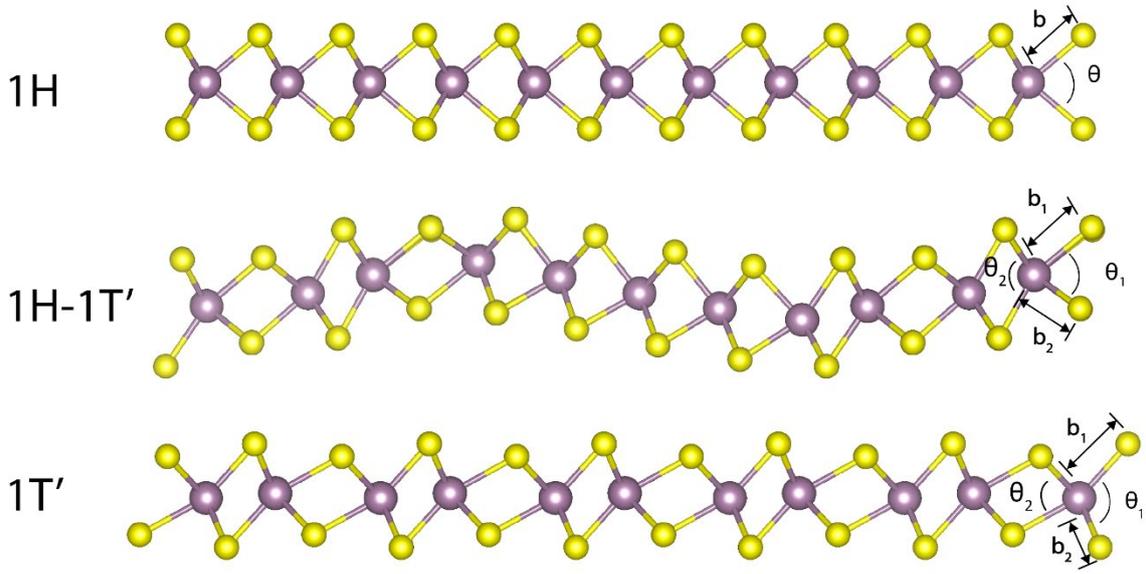

**Figure 1**. Side view of the optimized atomic structures of 1H, 1H/1T´, and 1T´ TMD. Color code: yellow, S/Se, purple: Mo/W.

**Table 1**. Calculated Lattice Constants $a$ (Å) and $b$ (Å), Monolayer Thickness $t$ (Å), Bond Length $d$ (Å), and X−TM−X Bond Angles $\theta$ (degree) of monolayer TMDs

| TMD | 2D crystal system | $a$ (Å) | $b$ (Å) | $t$ (Å) | $d$ (Å) | $\theta$ (deg) |
|---|---|---|---|---|---|---|
| $MoS_2$ | 1H - Hexagonal | 3.17 (3.15[31]) | 3.17 | 3.13 | 2.41 | 81.5 |
|  | 1H/1T'- Oblique | 25.59 | 12.65 | 4.69 | 2.36-2.55 | 79.9, 83.0 |
|  | 1T'- Oblique | 6.32 | 6.51 | 3.47 | 2.38-2.52 | 80.3, 85.8 |
| $MoSe_2$ | 1H - Hexagonal | 3.29(3.29[32]) | 3.29 | 3.35 | 2.54 | 82.9 |
|  | 1H/1T'- Oblique | 26.63 | 13.08 | 4.96 | 2.49 - 2.68 | 76.3, 81.3 |
|  | 1T'- Oblique | 6.52 | 6.78 | 3.76 | 2.51 - 2.64 | 80.3, 84.2 |
| $WS_2$ | 1H - Hexagonal | 3.18(3.18[33]) | 3.18 | 3.15 | 2.42 | 81.5 |
|  | 1H/1T'- Oblique | 25.66 | 12.70 | 4.72 | 2.39 - 2.57 | 78.4, 80.3 |
|  | 1T'- Oblique | 6.37 | 6.53 | 3.49 | 2.40 - 2.51 | 81.0, 85.6 |
| $WSe_2$ | 1H - Hexagonal | 3.29(3.26[32]) | 3.29 | 3.38 | 2.54 | 83.3 |
|  | 1H/1T'- Oblique | 26.62 | 13.08 | 5.01 | 2.52 - 2.69 | 76.1, 82.0 |
|  | 1T'- Oblique | 6.55 | 6.77 | 3.80 | 2.52 - 2.65 | 80.4, 83.9 |

The calculated lattice constants, monolayer thicknesses, and bond lengths are listed in **Table 1**. The 1H phase has the highest symmetry with the smallest unit cell. The corresponding lattice constant of $MoS_2$, $MoSe_2$, $WS_2$ and $WSe_2$ monolayers is 3.16, 3.29, 3.17 and 3.29 Å, respectively, which are close to the reported values [34-37]. For the 1T' phase, the lattice constants of *a* and *b* are different, which increases the anisotropy of the systems due to the reduced symmetry from the hexagonal to the oblique. In the lateral 1H/1T' heterostructure, the anisotropic features become more noticeable, which lead to the much large monolayer thickness. In our model, there are 4 $MoS_2$ units with the edge-sharing $TMX_6$ trigonal prisms structure and 4 more $MoS_2$ units with the edge-sharing $TMX_6$ octahedra, which are analogue to 1H and 1T' phases, respectively, along the x direction. As evidenced by Table 1, the lattice constants of three phases are similar to each other when the X anions are same. It suggests the lattice constants of the three phases are mainly determined by the X anions. This is ascribed to the different radius of ions. The $Mo^{4+}$ and $W^{4+}$ have the similar radius of 0.79 and 0.80 Å, respectively. As a comparison, the $S^{2-}$ of 1.84 Å is 0.14 Å smaller than that of $Se^{2-}$.

The electrical conductivity of different phases of TMDs was investigated through the analyses of their total density of states (TDOS), as shown in **Figs. 2-5**. It can be found that all 1H TMD monolayers are semiconductors. The bang gap energies of $MoS_2$, $MoSe_2$, $WS_2$ and $WSe_2$ in their 1H structural phases are 1.75, 1.51, 1.85 and 1.64 eV, respectively. Our calculated band gap at the PBE-GGA level of 1H-$MoS_2$ well match the experimental value (1.8 eV)[38]. While the PBE results often underestimate the band gap energy of semiconductor[39], our results are agreement with the reported values at the same functional level[40]. It reveals that the band gaps increase by the increase of the atomic number of TM cations. However, the band gaps decrease by the increase of the X anions. Such variation can be ascribed the electronegativity of atoms. According to the definition and trend of the atomic electronegativity in the periodic table, the metal atoms become more active with the increase of the atomic number. And the non-metal atoms have the opposite trend. Consequently, the larger TM with the smaller X have the stronger cohesive energies in TMD, which leads to the large band gap energies. As a comparison, all the 1T' monolayers are metallic. Some recent studies reveal that the 1T' $MoS_2$ may have a small band gap [41, 42]. Our results reveal that the evolution at Fermi energy level is very small. Such subtle difference may be ascribed to the XC functional used here. In the lateral 1H/1T' heterostructures, the PDOS of Mo and S atoms in the 1H and 1T' areas were separated calculated. It demonstrates that the Mo and S atoms in the 1T' part keep the metallic characteristics. In the 1H part, the S and Se p states are similar to that in the 1H monolayers. However, the Mo and W d states change slightly with the slight metallic features after the hybridization of the 1T' and 1H phases. Moreover, a big conduction band peak next to Fermi energy level the It suggests that the formation of the lateral heterostructures has considerable influence on the electronic properties of TM cations.

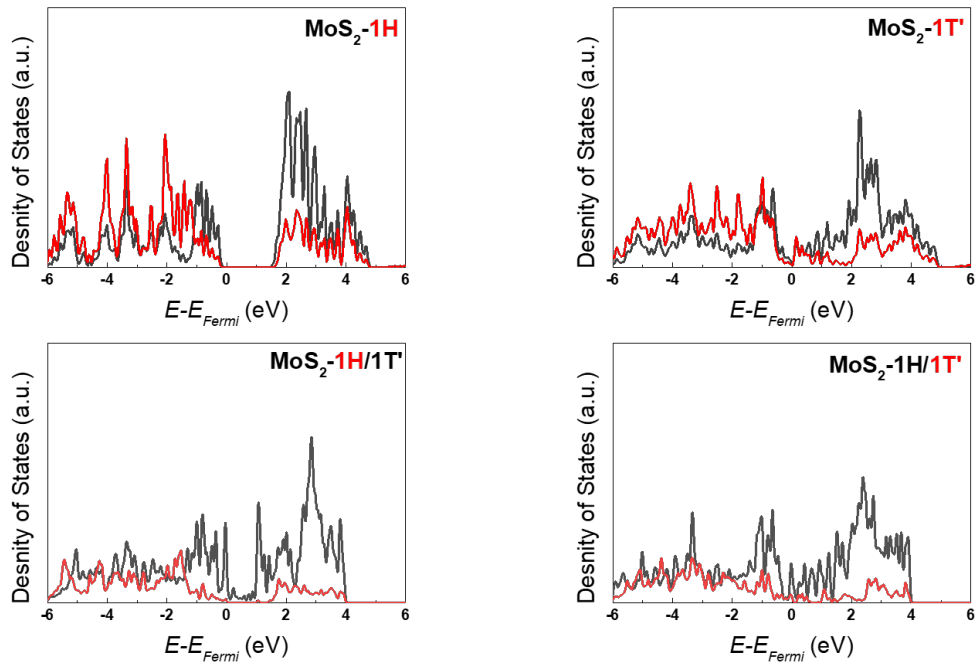

**Figure 2**. Partial DOS of the Mo 4d (grey line) and S 3p (red line) states of $MoS_2$ in the 1H monolayer, 1T' monolayer, 1H area of 1H/1T' monolayer and 1T' area of 1H/1T' monolayer.

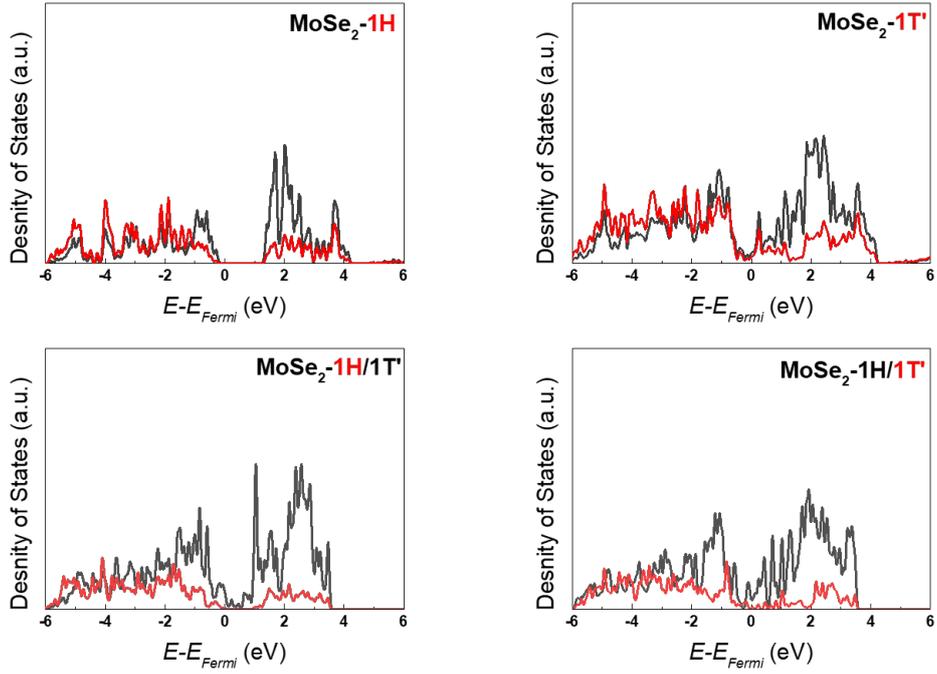

**Figure 3.** Partial DOS of the Mo 4d (grey line) and Se 4p (red line) states of MoSe$_2$ in the 1H monolayer, 1T' monolayer, 1H area of 1H/1T' monolayer and 1T' area of 1H/1T' monolayer.

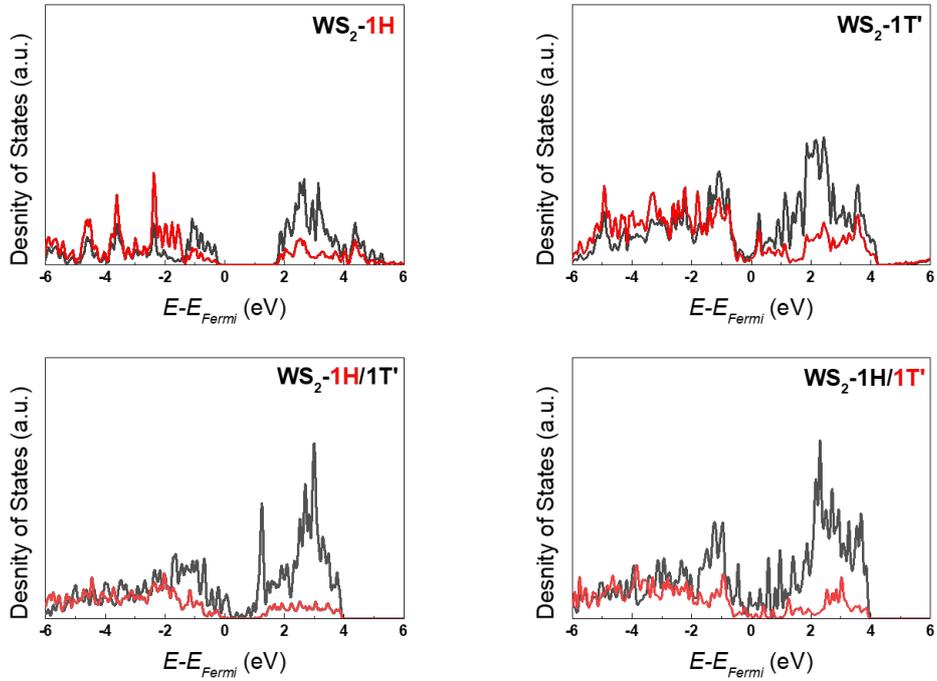

**Figure 4.** Partial DOS of the W 5d (grey line) and S 3p (red line) states of WS$_2$ in the 1H monolayer, 1T' monolayer, 1H area of 1H/1T' monolayer and 1T' area of 1H/1T' monolayer.

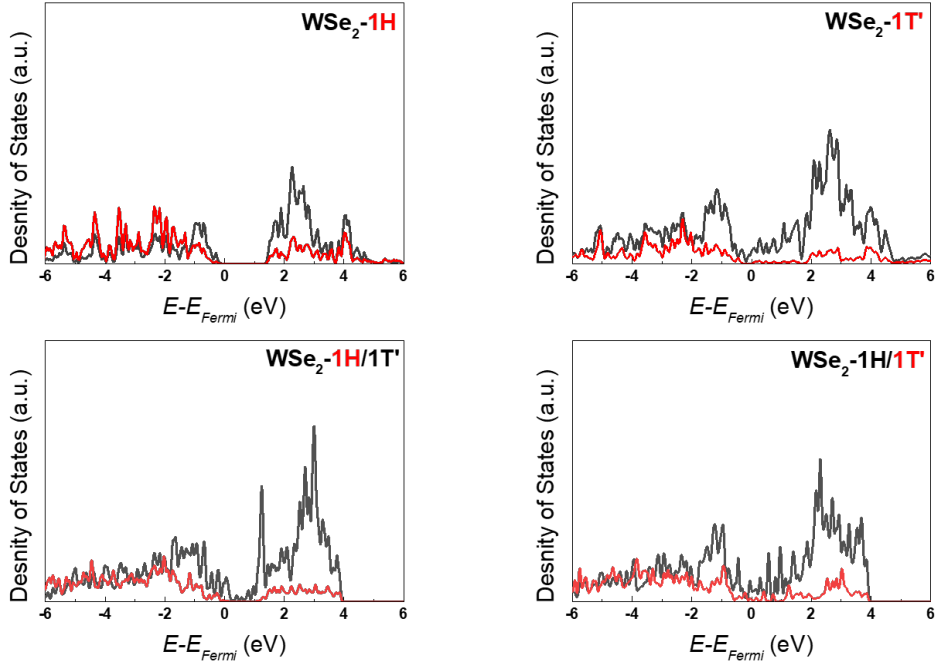

**Figure 5**. Partial DOS of the W 5d (grey line) and Se 4p (red line) states of $WSe_2$ in the 1H monolayer, 1T' monolayer, 1H area of 1H/1T' monolayer and 1T' area of 1H/1T' monolayer.

The calculated in-plane elastic constants for all TMDs with different structural phases are listed in **Table 2**. According to the Born-Huang criteria [43], our results demonstrate that all four phases of TMDs are mechanically stable. The mechanical properties of TMDs are also illustrated in **Fig. 6**, where it shows the impact of the structural phase on the mechanical characteristics of the 2D materials. The in-plane stiffness tensors $C_{11}$, $C_{22}$ and $C_{12}$ are widely used parameters of the TMDs for the comparison of their mechanical strength. It can be found that the $C_{11}$ and $C_{22}$ values are much larger than that of other in-plane stiffness tensors. Meanwhile, $C_{11}$ and $C_{22}$ are also greatly affected by the change of the structural phase. As evidenced by **Table 2**, our calculated in-plane stiffness tensors of TMD with their 1H phases are close to the reported values, which validates the method used for this study. The mechanical properties of TMD in their 1H phases are strongest. The strong 1H phase indicates the TM-X bonding strength is larger, which can be ascribed their different electronic configurations. In the trigonal prism of the 1H phase, the TM d orbitals are splitting into three levels: $d_{z2}$, $d_{x2-y2}$ + $d_{xy}$, and $d_{xz}$ +$d_{yz}$. The two valence electrons of TM stay in the lowest $d_{z2}$ level in 1H phase, which leads to strong valence bonding strength and semiconductor characteristics (see Fig. 2-5). As a comparison, 1T' has a weaker bond strength because two valence electrons stay at the $t_{2g}$ orbitals in the octahedra crystal field. The different bonding strength can be supported by the shorter average TM-X bond length in the 1H phase. The maximum $C_{11}$ value belongs to $WS_2$ in their 1H structural phase at 148 N m$^{-1}$, which is close to the reported value of 144 N m$^{-1}$[44]. And the 1H $MoSe_2$ has the smallest $C_{11}$ value of 113 N m$^{-1}$ in comparison with the other 1H TMDs. Interestingly, the trend of the $C_{11}/C_{22}$ in terms of the TM and X ions is same as that of the band gap energy in their 1H phase. It supports that the mechanical properties are determined by the cohesive energies of TMD, which is determined by the electronegativity trend of atoms. However, the $C_{12}$ values, which are related to the strain along the *xy* directions, almost same with the change of the TM cation ions. And the $C_{12}$ of $TMSe_2$ will slightly decrease in comparison with that of $TMS_2$. The change of the $C_{11}$ and $C_{22}$ caused by the X anions are also larger than that by TM cations. It suggests that the effect of the anions on the mechanical properties of TMD is larger.

For the 1T' phase, The $C_{11}$ and $C_{22}$ values of all TMD monolayers considered here decreases in comparison with that with the 1H phase. Again, it matches the previous study that the 1H phase is more thermodynamically stable caused by the large cohesive energies. The similar trend in terms of the TM and X ions are observed. Additionally, the $C_{11}$ and $C_{22}$ are inequivalent due to the decreased symmetry from the hexagonal (1H) to the oblique (1T').

**Table 2.** Calculated elastic constants $C_{11}$, $C_{22}$ and $C_{12}$ (N m$^{-1}$) of TMD monolayers in different structural phases. The data in the brackets are the reported results [44].

| TMD | Structural phase | $C_{11}$ (N m$^{-1}$) | $C_{22}$ (N m$^{-1}$) | $C_{12}$ (N m$^{-1}$) | $C_{16}$ (N m$^{-1}$) | $C_{26}$ (N m$^{-1}$) | $C_{66}$ (N m$^{-1}$) |
|---|---|---|---|---|---|---|---|
| MoS$_2$  | 1H | 134 (130) | 134 (130) | 29 (32) | 0 | 0 | 52 |
|          | 1H/1T' | 90 | 59 | -5 | 9 | 5 | 38 |
|          | 1T' | 108 | 119 | 21 | 0 | 0 | 41 |
| MoSe$_2$ | 1H | 113 (108) | 113 (108) | 24 (25) | 0 | 0 | 44 |
|          | 1H/1T' | 91 | 78 | 5 | 2 | 4 | 36 |
|          | 1T' | 92 | 111 | 29 | 0 | 0 | 40 |
| WS$_2$   | 1H | 148 (144) | 148 (144) | 32 (31) | 0 | 0 | 58 |
|          | 1H/1T' | 102 | 101 | 28 | -5 | 4 | 38 |
|          | 1T' | 118 | 139 | 22 | 0 | 0 | 49 |
| WSe$_2$  | 1H | 124 (119) | 124 (119) | 21 (22) | 0 | 0 | 52 |
|          | 1H/1T' | 98 | 101 | 11 | 1 | -3 | 48 |
|          | 1T' | 115 | 131 | 28 | 0 | 0 | 46 |

The TMD monolayers with the lateral 1H/1T' heterostructure have the smallest $C_{11}$ and $C_{22}$ values, which indicates that they possess the weakest mechanical strength and least thermodynamic stability. The low elastic constants suggest the transition from 1H to 1T' weakens the TM-X bonding strength, which can be ascribed to the dislocation at the interface due to the phase change from the edge-sharing TMX$_6$ trigonal prisms 1H phase to edge-sharing TMX$_6$ octahedra 1T' phase. The weak TM-X bonding strength in the heterostructures can be supported by some relatively larger TM-X bond lengths, as listed in Table 1. As such, the interface may be easier to deform with high strain, which can further lead to the formation of defects. Same as the other two phases, the WS$_2$ is the strongest one with the largest $C_{11}$ and $C_{22}$ value of 102 and 101 N m$^{-1}$, respectively. And the MoSe$_2$ is the weakest one. Interestingly, the $C_{12}$ values show the different trend in terms of the structural phases for TMDs with the different X anions. For TMS$_2$, the $C_{12}$ value of the 1H is the largest, followed by 1H/1T' heterostructures. And the 1T' has the smallest $C_{12}$ value. However, for TMSe$_2$, the 1H/1T' heterostructures have the smallest $C_{12}$ values while the 1H phase still possesses the largest $C_{12}$. Again, it supports that the X anions have the greater impact on the mechanical properties. This also matches the trend of the band gap. The difference between TMDs with the different TM cation is less than 0.15 eV. However, such difference with the different X anions is more than 0.20 eV.

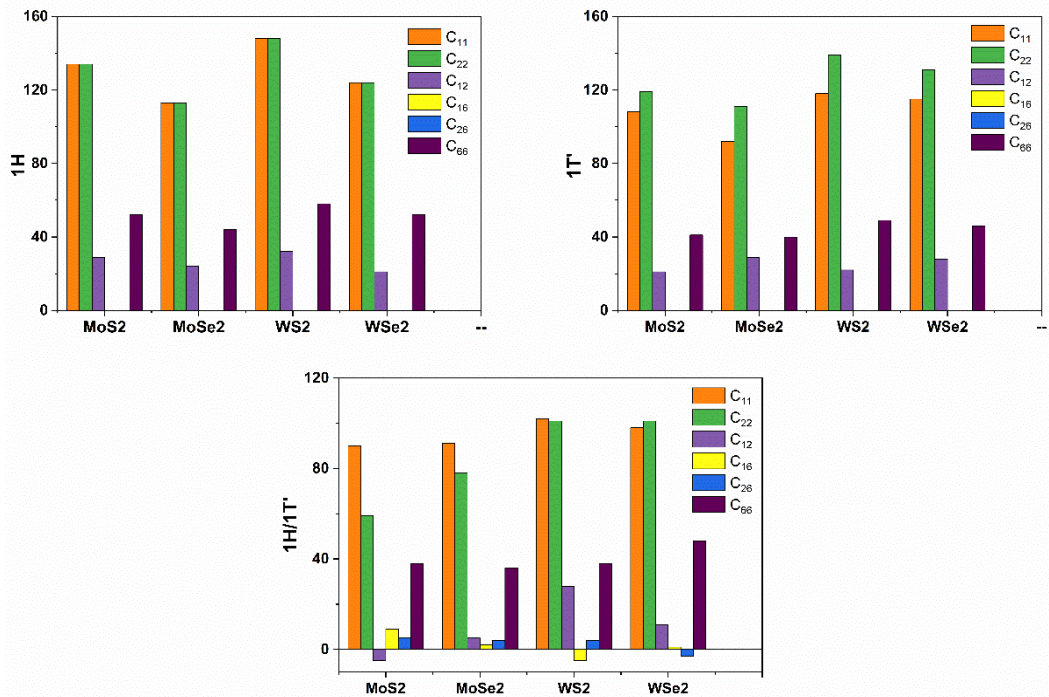

**Figure 6**. In-plane stiffness tensors $C_{ij}$ of TMD structures with different structural phases.

## IV. CONCLUSIONS

In summary, the structural, electronic and mechanical properties of MoS$_2$, WS$_2$, MoSe$_2$, and WSe$_2$ monolayers with the 1H, 1T' and 1H/1T' structural phases were investigated using the first principles DFT method. The in-plane stiffness tensors of these TMD monolayers were calculated to understand the behavior of the TMDs in their different phases. The comparative analyses on the electronic and mechanical properties also suggest that WS$_2$ is the most thermodynamically and mechanically stable materials in four TMDs studied here when they have the same structural phase. As a comparison, MoSe$_2$ is the weakest one. Our results further reveal that both TM cations and X anions can affect the electronic and mechanical properties. However, the impact of the X anion is relatively greater. Additionally, the structural phase transition can change the mechanical properties of the TMDs. Amongst the proposed phases, the monolayer TMDs in their 1H phase are the most mechanically stable materials, which make the 1H/1T' TMD lateral heterostructures the least mechanically stable structures. As such, when the lateral heterostructures are purposely designed for their application in catalysis, electronics and renewable energy generation areas, their mechanical stability needs to be carefully considered. On the hand, the relatively weak stiffness of the 1H/1T' heterostructure may be beneficial to the strain engineering of such materials. It is also worth noting that the atomic model of the 1H/1T' lateral heterostructure is still relatively small. The change of the size of the 1H and 1T' parts may significantly alter the overall mechanical properties of the lateral heterostructures, which deserves further investigation in the future. The findings of this study have offered theoretical guidance to manipulate the mechanical properties of TMD monolayers by tuning their structural phases.

**SUPPLEMENTAL MATERIAL**
Lattice constants and coordinate of 1H, 1T´ and 1H/1T' of MoS$_2$, MoSe$_2$, WS$_2$ and WSe$_2$.

**ACKNOWLEDGMENTS**
The authors thank Prof. Vei Wang for his insightful suggestions. This research was undertaken on the supercomputers in National Computational Infrastructure (NCI) in Canberra, Australia, which is supported by the Australian Commonwealth Government, and Pawsey Supercomputing Centre in Perth with the funding from the Australian government and the Government of Western Australia.

# REFERRENCES